\documentclass[prl,twocolumn,showpacs,preprintnumbers,amsmath,amssymb]{revtex4}

\usepackage{amsfonts} 

\usepackage{hyperref}

\usepackage{color} 
 
\usepackage[utf8,applemac]{inputenc} 
\newcommand{\bea}{\begin{eqnarray}}
\newcommand{\eea}{\end{eqnarray}}
\newcommand{\be}{\begin{equation}}
\newcommand{\ee}{\end{equation}}

\def\p{\partial}

\usepackage{mathtools}

\begin{document}
\title{Self-Consistent Adiabatic Inspiral and Transition Motion
}
  
\author{Geoffrey Comp\`ere$^\clubsuit$\footnote{geoffrey.compere@ulb.be} and Lorenzo K\"uchler$^\diamondsuit$\footnote{lorenzo.kuchler@ulb.be}}
\affiliation{\vspace{4pt}
$^{\clubsuit\diamondsuit}$ Universit\'e Libre de Bruxelles and International Solvay Institutes, C.P. 231, B-1050 Bruxelles, Belgium \\ $^\diamondsuit$ Institute for Theoretical Physics, KU Leuven, Celestijnenlaan 200D, B-3001 Leuven, Belgium
}

\begin{abstract}
The transition motion of a point particle around the last stable orbit of Kerr is described at leading order in the transition-timescale expansion. Taking systematically into account all self-force effects, we prove that the transition motion is still described by the Painlev\'e transcendent equation of the first kind. Using an asymptotically matched expansions scheme, we consistently match the quasi-circular adiabatic inspiral with the transition motion. The matching requires us to take into account the secular change of angular velocity due to radiation reaction during the adiabatic inspiral. 
\end{abstract}

\pacs{04.30.-w, 04.25.-g, 04.25.Nx, 11.10.Jj}

\maketitle

Binary coalescences are the loudest signals of all current and prospective gravitational wave observatories \cite{Abbott:2016blz,Audley:2017drz,Abbott:2020niy,Maggiore:2019uih}. Current waveform models of such events are sophisticated interpolations between results from numerical relativity, effective one-body (EOB) methods, the post-Newtonian--post-Minkowskian formalism and black hole perturbation theory, see Table III of \cite{Abbott:2020niy} for a list of references. The current and future high precision tests of General Relativity \cite{Abbott:2020jks} strongly motivate the understanding of possible systematic errors in current waveform models. 

In particular, the modeling of the transition from inspiral to merger of binaries is notoriously difficult since it occurs in the strong field regime. In the quasi-circular approximation, %\cite{Teukolsky:1974yv,Poisson:2004cw} 
nonperturbative resummation techniques have been used to obtain explicit models for comparable mass binaries \cite{Buonanno:2000ef,Buonanno:2005xu,Damour:2007xr,Damour:2009kr,Pan:2013rra} and small mass ratio binaries \cite{Nagar:2006xv,Bernuzzi:2010ty,Bernuzzi:2010xj,Bernuzzi:2011aj}. The deviation from quasi-circularity was estimated to be numerically small,  $d\log r/d\phi \lesssim 0.05$, even around the innermost stable circular orbit (ISCO) \cite{Buonanno:2000ef}. In the small mass ratio expansion, the transition regime was defined in the quasi-circular approximation and neglecting self-force effects \cite{Ori:2000zn} but it was later shown that non-quasi-circular corrections occur at the same order in this transition regime \cite{Kesden:2011ma}. For extensions, see  \cite{Sundararajan:2008bw,Taracchini:2014zpa,Apte:2019txp,Compere:2019cqe,Burke:2019yek}. 

The main aim of this Letter is to provide an accurate and complete treatment of the matching between the inspiral and transition motion in the small mass ratio regime, taking into account all self-force effects, with the motivation to extend current self-force models and provide more faithful EOB models. The inspiral motion can be studied via the slow timescale expansion \cite{Hinderer:2008dm}, which breaks at the separatrix between bound and plunging orbits \cite{Glampedakis:2002ya} or during resonances \cite{Flanagan:2010cd,Flanagan:2012kg}. Radiation reaction requires an inspiral with dynamical angular velocity. We will first derive such an inspiral in the adiabatic regime around a Kerr black hole, thereby extending the current quasi-circular parametrization with geodesic circular angular velocity \cite{Miller:2020bft,Pound:2021qin} to dynamical angular velocity. 

In the following, we will review the equations of equatorial forced geodesics, solve them in the adiabatic approximation for nongeodesic angular velocity and expand the solution close to the last stable orbit. We will then solve the equations again but in the transition-timescale expansion. We will finally match the two expansions in the overlapping region using the method of asymptotically matched expansions and conclude. 

\vspace{2pt} \noindent {\bf Conventions:} We use geometrical units $G=c=1$. All quantities are made dimensionless using the mass $M$ of the Kerr background, including the Kerr angular momentum $a$, the binary mass ratio $\eta=m/M$ where $m$ is the point-particle mass, the proper particle energy $e = -p_{t}/m$ and the proper particle azimuthal angular momentum $\ell = p_\phi/(mM)$.  Spacetime indices are lowered and raised with the Kerr metric in Boyer-Linquist coordinates $g_{\mu\nu}$. The outer horizon is the largest root of $\Delta=r^2-2r+a^2$.

\section{Equatorial forced geodesics}

We consider equatorial orbits around the Kerr black hole with position 
$z^\mu= ( t,  r, \frac{\pi}{2}, \int \Omega d t )$ where $\Omega=d\phi/d t$ is the orbital frequency. We denote as $\sigma=\text{sign}(\Omega)$, i.e., $\sigma=+1$ for prograde orbits and $\sigma=-1$ for retrograde orbits. The velocity is $v^\mu=dz^\mu/d\tau$ and, in particular, the redshift is denoted as  $U=dt/d\tau$ where $\tau$ is the dimensionless proper time. In terms of the angular momentum $\ell =v_\phi$ and energy $e=-v_t$ one has 
\begin{equation}\label{UO}
    U = -g^{tt}e+g^{t\phi}\ell,\quad \Omega=\frac{-g^{t\phi}e+g^{\phi\phi}\ell}{-g^{tt}e+g^{t\phi}\ell}. 
\end{equation}
It will be convenient to introduce $\delta$ as the deviation from the geodesic angular velocity $\Omega_{\text{geo}} = \sigma/(r^{3/2}+\sigma a)$ as  
\begin{equation}\label{valdelta}
    \Omega = \frac{\sigma}{r^{3/2}+\sigma a + \delta}  ,\qquad \delta =\sigma (\Omega^{-1}-\Omega_{\text{geo}}^{-1}).   
\end{equation}
The forced geodesic equations $v^\alpha \nabla_\alpha v^\mu=f^\mu$ and the normalization of the velocity $v^\mu v_\mu = -1$ are equivalent to (i) the radial equations
\begin{subequations}\label{kerreomeqr}
\begin{align}
    \left(\frac{d r}{d\tau}\right)^2 &=e^2- V^{\text{geo}}, \qquad \frac{d^2 r}{d \tau^2}+  \frac{1}{2} \frac{\p V^{\text{geo}}}{\p  r} = f^{ r},\\ 
    V^{\text{geo}} & \equiv 1-\frac{2}{r}+\frac{\ell^2+a^2(1-e^2)}{r^2}+\frac{2(\ell-a e)}{r^3};\label{Vgeo}
\end{align}
\end{subequations}
(ii) the energy and angular momentum flux-balance equations 
\begin{equation}
\frac{d\ell}{d\tau} =  f_\phi, \qquad  \frac{de}{d\tau} =-  f_{ t}; \label{fluxeq}
\end{equation}
and, (iii) the orthogonality of the force with the velocity, $f_{\mu} v^\mu =0$,  which can be written as
\begin{equation}
\label{defy}
    \Omega^{-1}\frac{de}{d \tau} - \frac{d\ell}{d \tau}= f_{r} \frac{d r}{d\tau}(\Omega U)^{-1}.
\end{equation}
%In the absence of radiation reaction forces,  $\frac{dr}{d\tau}=0$, this is the first law of binary mechanics at fixed masses \cite{LeTiec:2011ab}.

\section{Slow timescale expansion}
We now restrict our analysis to the small mass ratio limit $\eta \ll 1$ and to orbits without eccentricity. Such orbits can be described in the inspiral phase using the slow timescale expansion 
\begin{align}\label{inspiralexp}
 X&= X_{(0)}(\tilde \tau) + O_{\tilde \tau}(\eta)
\end{align}
where all variables are collectively denoted as  $X=(a,\delta,r,\Omega,U,e,\ell)$. Here and below, the indices in parentheses $(i)$ label the terms appearing at order $\eta^i$ in the expansion. The slow proper time is defined as $\tilde \tau \equiv \eta\,  \tau$. Since all quantities have been made dimensionless using a rescaling with the mass $M$, we will disregard the slow time evolution of $M$. The symbol $O_{\tilde \tau}(\eta)$ refers to the limit $\eta\rightarrow0$ at fixed slow proper time $\tilde \tau$. Neglecting such corrections defines the adiabatic approximation. 

Equations \eqref{fluxeq} and \eqref{defy} then lead to the expansion 
\begin{align}
f^{a} &= \eta \, f^{a}_{(1)}(\tilde \tau) + O_{\tilde\tau}(\eta^2) , \quad a=t,\phi,\label{expft}\\ 
f^r &= f_{(0)}^{r}(\tilde \tau)+ \eta \, f^{r}_{(1)}(\tilde \tau) + O_{\tilde\tau}(\eta^2) . \label{frin}
\end{align}
As we will discuss below, the consistent matching of the adiabatic inspiral with the transition motion will be consistent with canceling the leading-order radial self-force $f_{(0)}^{r}(\tilde \tau)$, see Eq. \eqref{finalfr} below. 

\subsection{Quasi-circular adiabatic inspiral}

The adiabatic solution without eccentricity to Eqs. \eqref{UO}--\eqref{kerreomeqr}--\eqref{fluxeq}--\eqref{defy} can be found straightforwardly. In order to write compact expressions, it is convenient to define the coefficients
\begin{align}
    A&=   r_{(0)}^3 - 3  r_{(0)}^2 + 2\sigma   a_{(0)}  r_{(0)}^{3/2},\nonumber\\
    B &=    r_{(0)}^2 - 2 \sigma   a_{(0)}   r_{(0)}^{1/2} +   a_{(0)}^2,\\
    C&=   r_{(0)}^{3/2} - 2   r^{1/2}_{(0)} + \sigma   a_{(0)},   \qquad 
    D= B (4A r_{(0)}^{-1}-3  \Delta),\nonumber
\end{align}
and their $\delta_{(0)}$-corrected version, 
\begin{subequations}\label{ABCDdelta}
\begin{align}\label{Ad}   
    A_\delta&=A+2 C \delta_{(0)} +(1-2 r_{(0)}^{-1}) \delta_{(0)}^2,    \\ \label{Bd}
    B_\delta &=  B-2\sigma   a_{(0)} r_{(0)}^{-1} \delta_{(0)} ,\\ \label{Cd}
    C_\delta &=C+(1-2 r_{(0)}^{-1}) \delta_{(0)},  \\ \label{Dd}
    D_\delta&=4A_\delta [B r_{(0)}^{-1}+\delta_{(0)}   r_{(0)}^{-1/2}(1+  a_{(0)}^2r_{(0)}^{-2})] \nonumber \\ 
&    -  \Delta [3 B_\delta +2\sigma\delta_{(0)}(2+  r_{(0)}^{-3/2}\delta_{(0)})\Omega^{-1}_{(0)}  r^{-1}_{(0)}]. 
\end{align}
\end{subequations}
Importantly, the function $D$ admits a single root outside the horizon at the location of the geodesic ISCO $r_*$, 
\begin{equation}
\frac{D}{B}\vert_* = r_{(0)*}^2-6r_{(0)*}+8 \sigma a_{(0)}\sqrt{r_{(0)*}}-3a_{(0)}^2= 0. \label{defrs}
\end{equation}
%\begin{equation}
%\frac{D}{B}\vert_* = r_*^2-6r_*+8 \sigma a\sqrt{r_*}-3a^2= 0. %\label{defrs}
%\end{equation}
The unique solution to Eqs. \eqref{UO}--\eqref{kerreomeqr} can be written as 
\begin{subequations}\label{all0}
\begin{align}\label{Oms}
\Omega_{(0)}(\tilde \tau) &=  \sigma (   r_{(0)}^{3/2}+\sigma   a_{(0)} +\delta_{(0)})^{-1}, \\ \label{Us}
U_{(0)} (\tilde \tau)  & = \sigma A_\delta^{-1/2} \Omega_{(0)}^{-1}= A_\delta^{-1/2} \vert\Omega_{(0)}\vert^{-1},\\
\label{lcirceqd}
\ell_{(0)}(\tilde \tau)   &= \sigma B_\delta A_\delta^{-1/2},    \qquad e_{(0)}(\tilde \tau) =  C_\delta A_\delta^{-1/2}. 
\end{align}
\end{subequations}
The radial self-force is algebraically determined as \footnote{We expect $f_{(0)}^r(\tilde\tau) \rightarrow 0$ as $\tilde\tau \rightarrow -\infty$. We also expect  $f_{(0)}^r(\tilde\tau) \rightarrow 0$ as $\tilde\tau \rightarrow 0 $ so that $f^r_{(0)}\rightarrow 0$ in the limit $\eta \rightarrow 0$.}
\begin{equation}\label{deffr}
f^{  r}_{(0)}(r_{(0)},\delta_{(0)},a_{(0)})=  \Delta U_{(0)}^2 \Omega_{(0)}^2 \frac{\delta_{(0)}}{r_{(0)}^4} (\delta_{(0)}  + 2r_{(0)}^{3/2} )    .
\end{equation}
Equation \eqref{defy} is then equivalent to 
\begin{equation}\label{da}
\frac{d  a_{(0)}}{d\tilde \tau} = 0, 
\end{equation}
which implies that $  a_{(0)}$ is a constant. The flux-balance equations \eqref{fluxeq} are equivalent to
\begin{eqnarray}\label{r0inspiral}
\frac{d  r_{(0)}}{d\tilde \tau}  &=& \frac{r_{(0)}^2 A_\delta}{\delta_{(0)} (\delta_{(0)}+2 r_{(0)}^{3/2})}\left(e_{(0)} f_{(1)}^{  t}-\ell_{(0)} f_{(1)}^\phi \right) , \\\label{dinspiral}
\frac{d\delta_{(0)}}{d\tilde \tau} &=& -\frac{A_\delta^{3/2}}{B_\delta}f^{  t}_{(1)} + \frac{\sqrt{r_{(0)}}D_\delta}{2  \Delta B_\delta} \frac{d  r_{(0)}}{d\tilde \tau}. 
\end{eqnarray}
These equations are linear in the first order self-force and nonlinear in the kinematic parameters $  r_{(0)}(\tilde \tau)$, $\delta_{(0)}(\tilde \tau)$ and $  a_{(0)}$. Since the self-force is an integral over the past motion of the source, these evolution equations are retarded integro-differential equations. The adiabatic inspiral is quasi-circular in the sense that 
\begin{equation}
    \frac{d\log r}{d\phi}=\frac{dr/d\tilde t}{r d\phi/d\tilde t}=\frac{\eta}{r \Omega U}\frac{dr}{d\tilde\tau}=O_{\tilde \tau}(\eta). \label{QC1}
\end{equation}

\subsection{Inspiral towards the last stable orbit}

As demonstrated in \cite{Ori:2000zn,Kesden:2011ma}, the slow timescale expansion breaks at the ISCO in the absence of radial self-force corrections, i.e., $\delta\equiv 0 \equiv \delta_{(0)}$. This breakdown is clear from Eq. \eqref{dinspiral} since the left-hand side is now zero, $f^t_{(1)}\neq 0$ at the ISCO while we have $D_\delta =D=0$, which implies that $dr_{(0)}/d\tilde \tau$ blows up. Taking into account radial self-force corrections, it is physically expected that there will still be a breakdown of the evolution equations \eqref{r0inspiral} and \eqref{dinspiral}. We now assume that there exists a finite $r_{(0)*}$ and corresponding slow proper time $\tilde \tau_*$ where this breakdown occurs. Moreover, we assume that $d\delta_{(0)}/d\tilde \tau$ does not blow up at $\tilde \tau=\tilde\tau_*$. These two hypotheses will be justified \emph{a posteriori} by the consistent matching of the inspiral with the transition motion which implies, in particular, the match of Eq. \eqref{delta0in} with Eq. \eqref{delta0tr}. We will now prove under these two assumptions that $r_{(0)}=r_{(0)*}$ is exactly the location of the geodesic ISCO.

None of the quantities $f^{a}_{(1)}$, $A_\delta$, $B_\delta$, $C_\delta$, $  \Delta$ can diverge along the trajectory. We deduce that $\delta_{(0)}$ vanishes at $\tilde \tau_*$, $\delta_{(0)*}\equiv \delta_{(0)}(\tilde \tau_*)=0$. Moreover, since $d\delta_{(0)}/d\tilde \tau$ is finite at $\tilde \tau =\tilde\tau_*$, Eq. \eqref{dinspiral} implies that $D_\delta=0$ at $\tilde \tau =\tilde\tau_*$ in order to cancel the divergence of $dr_{(0)}/d\tilde \tau$. Now $\delta_{(0)*}=0$ implies $D_\delta = D$ at $\tilde \tau =\tilde\tau_*$. We deduce from Eq. \eqref{defrs} that $r_{(0)*}$ is the location of the geodesic ISCO. Even in the presence of self-force, the radial potential is given by the geodesic potential \eqref{Vgeo}. The last stable orbit (LSO) is instead defined from the radius $r$ where 
\begin{equation}\label{iscocond}
    \left.\frac{\p^2 V^{\text{geo}}(e,\ell,r,a)}{\p   r^2}\right. = [D +O(\delta_{(0)})]+ O_{\tilde\tau}(\eta) = 0
\end{equation}
after using $e=e_{(0)}$, $\ell=\ell_{(0)}$ as given in Eq. \eqref{lcirceqd}. Here, $O(\delta_{(0)})$ denote $\eta$-independent terms at least linear in $\delta_{(0)}$. At leading order in the small mass ratio expansion, the LSO therefore coincides with the ISCO since $ \delta_{(0)*}=0= D \vert_*$. Hence, we can either use the terminology of LSO or ISCO at this order though we expect both concepts will differ once subleading corrections in the mass ratio are taken into account.

We denote as $e_{(0)*}=C/\sqrt{A}\vert_*$, $\ell_{(0)*}=\sigma B/\sqrt{A}\vert_*$, $\Omega_{(0)*}=\sigma/(r_{(0)*}^{3/2}+\sigma a_{(0)})$ the energy, angular momentum, and angular velocity at the ISCO in the adiabatic limit. We assume that the energy and angular momentum admit an expansion close to the LSO with half-integer powers of $(\tilde \tau_*-\tilde \tau)$. Since $f_{(1)}^{t}$, $f_{(1)}^{\phi}$ are finite at the LSO, we assume consistently with Eq. \eqref{fluxeq} the expansion:
\begin{eqnarray}\label{inspiralexplicit}
    \ell_{(0)} -\ell_{(0)*} &=&\kappa^*_{(0),2} (\tilde \tau_*-\tilde \tau) + \kappa^*_{(0),3} (\tilde \tau_*-\tilde \tau)^{3/2}\nonumber \\
    && + O(\tilde \tau_*-\tilde \tau)^{2} , \\
    e_{(0)}-e_{(0)*}&=&\Omega_{(0)*} \kappa_{(0),2}^*(\tilde \tau_*-\tilde \tau) + e_{(0),3}^* (\tilde \tau_*-\tilde \tau)^{3/2}\nonumber\\ 
    && + O(\tilde \tau_*-\tilde \tau)^{2} .
\end{eqnarray}
Here and below the indices after a comma $,i$ refer to terms appearing at order $(\tilde \tau_*-\tilde \tau)^{i/2}$. The inspiral motion then implies the following expansion 
\begin{eqnarray}\label{inspiralexplicit2}
\! \!     r_{(0)}-r_{(0)*}&\!\!=\!\!&  r^*_{(0),1} (\tilde \tau_*-\tilde \tau)^{1/2} + O(\tilde \tau_*-\tilde \tau) , \\
\!\!    \delta_{(0)} &\!\!=\!\!& \delta_{(0),2}^* (\tilde \tau_*-\tilde \tau) + O(\tilde \tau_*-\tilde \tau)^{3/2}, \label{delta0in}\\
\! \!       \Omega_{(0)} -\Omega_{(0)*}&\!=\!& \Omega_{(0),1}^*(\tilde \tau_*-\tilde \tau)^{1/2} + O(\tilde \tau_*-\tilde \tau) , \\
\! \!       f^r_{(0)} &\!\!=\!\!& \frac{8}{3}r_{(0)*}^{-7/2}\delta_{(0)} + O(\tilde \tau_*-\tilde \tau)^{3/2}, \label{frval}
\end{eqnarray}
after using $4A_*=3r_{(0)*}\Delta_*$. Solving Eqs. \eqref{fluxeq}, \eqref{all0}, \eqref{r0inspiral} and \eqref{dinspiral}, we obtain $\Omega_{(0),1}^* = -\frac{3}{2}\sigma r_{(0)*}^{1/2}\Omega_{(0)*}^2r^*_{(0),1}$ and 
%the following constraints
\begin{align}
    \delta_{(0),2}^*&=\frac{\Omega_{(0)*}}{\Delta_*} \left( \frac{3\sigma r_{(0)*} E_*}{2A_*} (r^*_{(0),1})^2 -A_*^{3/2}\kappa^*_{(0),2}\right), \label{cons2}\\
    e_{(0),3}^*&= \Omega_{(0)*} \bigg( \frac{3 E_* \Omega_{(0)*} r_{(0)*}^{3/2}}{4A_*^{5/2}} (r^*_{(0),1})^3 \nonumber \\
    & -\frac{3\sigma}{2} r_{(0)*}^{1/2}\Omega_{(0)*} r^*_{(0),1}\kappa^*_{(0),2}+\kappa^*_{(0),3}\bigg), \label{cons3}
\end{align}
where $E_*\!\equiv\!-13a_{(0)}^4+3a_{(0)}^4r_{(0)*}+a_{(0)}\sigma(107-11a_{(0)}^2)r_{(0)*}^{3/2}-3(44-7a_{(0)}^2)r_{(0)*}^2+25r_{(0)*}^3$. The free parameters  $\kappa^*_{(0),2}$, $\kappa^*_{(0),3}$, $r^*_{(0),1}$ will be fixed from the matching with the transition-timescale expansion to which we turn. 

\section{Transition-timescale expansion}

We consider the expansion in the transition timescale
\begin{equation}
\label{s}
s \equiv \eta^{1/5} \left( \tau -  \tau_*\right)
\end{equation}
around the LSO crossing time $\tau_*$ or $s_*=0$. The transition motion will be defined from $s=-\infty$ (where it will asymptotically match the inspiral) up to the merger time $s=s_{\text{merger}}>0$ after which the motion will lie behind the black hole horizon $r=1+\sqrt{1-a^2}$. The validity of the transition equations will be assessed in Eq. \eqref{rangetr} below.

We define the variables $R$, $\xi$ and $Y$ as \cite{Ori:2000zn,Kesden:2011ma}
\begin{subequations}\label{exptau}
\begin{align}
& r - r_{[0]*} = \eta^{2/5}R(\eta,s),\;\; \;
 \ell -\ell_{[0]*} = \eta^{4/5}\xi(\eta,s), \\ 
&    e -e_{[0]*} = \Omega_{[0]*} [\eta^{6/5}Y(\eta,s) + \eta^{4/5}\xi(\eta,s)],
\end{align}
\end{subequations}
where the LSO values are expanded in powers of $\eta^{1/5}$ as 
\begin{align}
    r\vert_* \!\!&= \!\!  r_{[0]*} \!\!+\!\! \eta^{2/5}\sum_{i=0}^\infty \eta^{i/5}  r_{[i]*},
    \;\;
    e\vert_* \!\!=\!\! e_{[0]*} \!\!+\!\! \eta^{4/5}\sum_{i=0}^\infty \eta^{i/5}  e_{[i]*},\nonumber
    \\
    \ell\vert_* \!\!&=\!\! \ell_{[0]*} \!\!+\!\! \eta^{4/5}\sum_{i=0}^\infty \eta^{i/5}  \ell_{[i]*},
    \;\;
    \Omega\vert_* \!\!=\!\! \Omega_{[0]*} \!\!+\!\! \eta^{2/5}\sum_{i=0}^\infty \eta^{i/5}\Omega_{[i]*}. \nonumber
\end{align}
The values $R\vert_*$, $\xi\vert_*$, $Y\vert_*$ encode the shifts of these quantities at the LSO, i.e., $R\vert_*=\sum_{i=0}^\infty   r_{[i]*}\eta^{i/5}$, \dots. The expansion of $\Omega\vert_*$ is consistent with Eq. \eqref{UO}. The indices in square brackets $[i]$ label the terms appearing at relative order $\eta^{i/5}$ with respect to the first nonvanishing leading term $[0]$ in the transition-timescale expansion.

In the absence of radial self-force all variables $R,\xi,Y,a$ scale as $\eta^0$ in the transition region for standard spins \cite{Ori:2000zn,Kesden:2011ma,Compere:2019cqe,Burke:2019yek}. In the presence of radial self-force, we will assume the same scaling and show consistency. We therefore expand
\begin{eqnarray}
R&=&\sum_{i=0}^\infty \eta^{i/5} R_{[i]}(s), \quad Y=\sum_{i=0}^\infty \eta^{i/5} Y_{[i]}(s), \label{expR} \\
\xi&=&\sum_{i=0}^\infty \eta^{i/5} \xi_{[i]}(s), \quad a = a_{[0]} + \sum_{i=1}^\infty \eta^{i/5} a_{[i]}(s).  
\end{eqnarray}
Consistently with Eqs. \eqref{fluxeq} and \eqref{defy} we have
\begin{eqnarray}
f_a &=&  f_{a[0]}(s) \eta +O_s(\eta^{6/5}), \quad a=t,\phi,\label{fa0}\\ 
f^r &=& f^r_{[0]}(s) \eta +O_s(\eta^{6/5}),\label{fr4}
\end{eqnarray}
where $O_s(\eta)$ refers to terms of order $\eta$ at fixed $s$. The angular momentum flux-balance law \eqref{fluxeq} becomes $\xi(s)=\xi_{[0]}(s)+O_s(\eta^{1/5})$ with $d\xi_{[0]}/ds=f_{\phi[0]}(s)$. 

Instead of Eq. \eqref{QC1}, the orbit is now quasi-circular in the weaker sense
\begin{equation}
    \frac{d\log r}{d\phi}=\frac{dr/dt}{r \Omega}=\frac{\eta^{3/5}}{r \Omega U}\frac{dR}{ds}=O_s(\eta^{3/5}).    \label{QC2}
\end{equation}

\subsection{Leading-order transition equations}

We now derive the solution to Eqs. \eqref{UO}, \eqref{kerreomeqr}--\eqref{defy} at leading order in the transition-timescale expansion around the LSO. The condition \eqref{iscocond} together with Eqs.  \eqref{UO}--\eqref{kerreomeqr} give at leading order in $\eta$ and at the LSO the same quantities $a_{[0]*}=  a_{(0)}$, $  r_{[0]*}=  r_{(0)*}$, $\delta=O_s(\eta^{4/5})$, $e_{[0]*} = e_{(0)*}$, $\ell_{[0]*} = \ell_{(0)*}$, $\Omega_{[0]*}=\Omega_{(0)*}$ as the adiabatic inspiral.

From Eq. \eqref{kerreomeqr} we obtain as algebraic equations $  a_{[1]}=  a_{[2]}=  a_{[3]}=0$, while $  a_{[4]}$, $a_{[5]}$, and $a_{[6]}$ are proportional to $D\vert_*$ and therefore vanish as well from Eq. \eqref{defrs}. This matches with the constancy of $a_{(0)}$ \eqref{da} in the adiabatic inspiral. Using Eqs. \eqref{UO} and \eqref{iscocond} the deviation $\delta$ defined in Eq. \eqref{valdelta} is given around the LSO at leading order as 
\begin{equation}
\delta = \delta_{[0]} \eta^{4/5} \!+\! O_s(\eta),\;\;  \delta_{[0]}=\frac{\pi_*}{\Delta_*^2}R_{[0]}^2(s)-\frac{A_*^{3/2} \Omega_{[0]*}}{\Delta_*} \xi_{[0]}(s) \label{delta0tr}
\end{equation} 
where $\pi_*\!\equiv\! a_{(0)}^3\sigma \!-\! 6 a_{(0)} \sigma r_{(0)*} \!+\! 2(3+a_{(0)}^2)r_{(0)*}^{3/2} \!-\! 3 a_{(0)}\sigma r_{(0)*}^2$. Any quantity $X(r,\delta,a)$ that is finite at the LSO can now be expanded as 
\vspace{-6pt}
\begin{align}\label{X}
X&=X_{[0]*} +\eta^{2/5}\left. \frac{\partial X}{\partial r}\right\vert_{[0]*} R(s)+\eta^{4/5} \left( \left. \frac{\partial X}{\partial \delta} \right\vert_{[0]*} \delta_{[0]}(s) \right. \nonumber \\
&\left. + \frac{1}{2} \left. \frac{\partial^2 X}{\partial r^2}\right\vert_{[0]*} R_{[0]}^2(s) \right) + O_s(\eta).
\end{align}
In particular for $\eta^{-1}f_\phi(r,\delta,a)$, comparing with Eq. \eqref{fa0} tells that $f_{\phi[0]}(s)=f_{\phi[0]}$ is a constant. We define 
\begin{equation}
    \kappa_* \equiv - f_{\phi[0]}\label{kappa}
\end{equation}
so that $\xi_{[0]}(s)\!\!=\!\!-\kappa_* s$. It is clear that $\kappa_*>0$ since angular momentum loss drives the transition motion. For $f^r(r,\delta,a)$, comparing with Eq. \eqref{fr4} leads to 
\begin{eqnarray}
&&\!\!\!\!f^r_{[0]}(s) =f^r_{[0]*},\;\;\;\; f^r_{[4]}(s) = \epsilon_* R_{[0]}^2(s)- \zeta_*\xi_{[0]}(s) ,\label{defepszeta} \\
&&\!\!\!\! \zeta_* \equiv  \frac{A_*^{3/2}\Omega_{[0]*}}{\Delta_*} \left. \frac{\partial f^r}{\partial \delta}\right\vert_{[0]*}\!\!\!\!, \;\;\; \epsilon_* \equiv  \left. \frac{1}{2}\frac{\partial^2 f^r}{\partial r^2}\right\vert_{[0]*} \!\!\!+\frac{\pi_*}{\Delta^2_*} \left.\frac{\partial f^r}{\partial \delta}\right\vert_{[0]*}\!\!\!\!,\nonumber
\end{eqnarray}
with $f_{[2]}^r$ and $f_{[3]}^r$ also non-vanishing. Expanding Eq.~\eqref{kerreomeqr}, the leading-order transition equations are then given by 
\begin{eqnarray}
\!\!\! &&   \left(\frac{d R_{[0]}}{ds}\right)^2 = -\frac{2}{3}\alpha_*R_{[0]}^3 - 2\beta_* \kappa_* s R_{[0]} + \gamma_*Y_{[0]}, \nonumber\\
\!\!\! &&   \frac{d^2R_{[0]}}{ds^2} = -\alpha_*R_{[0]}^2 -\kappa_*  \beta_*s, \\
\!\!\! &&    \frac{d Y_{[0]} }{ds}  \!=\!  2\kappa_*\frac{\beta_*}{\gamma_*}R_{[0]} \nonumber
\end{eqnarray}
where the coefficients read as
\begin{subequations}\label{defcoefs} \begin{align}
    \alpha_*& \equiv \frac{1}{4}\left.\frac{\p^3V^\text{geo}}{\p  r^3}\right\vert_{[0]*},\qquad \gamma_* \equiv \left.\frac{\p V^\text{geo}}{\p \ell}\right\vert_{[0]*},\\
    \beta_* &\equiv -\frac{1}{2}\left.\left(\frac{\p^2V^\text{geo}}{\p  r\p \ell} +  \Omega\frac{\p^2V^\text{geo}}{\p  r\p e}\right)\right\vert_{[0]*}.
\end{align} 
 \end{subequations}
Introducing
 \begin{subequations}\label{xytDef}
\begin{eqnarray}
\mbox{}\!\!\!\!\!\!  x_{[0]}  \!\!&\!\equiv \!\!\! &\alpha_*^{3/5}\beta_*^{-2/5}\kappa_*^{-2/5}R_{[0]},\\
\mbox{}\!\!\!\!\!\!  y_{[0]}\!\!&\!\equiv\!\!\! &\frac{\alpha_*^{4/5}\gamma_*}{\beta_*^{6/5}\kappa_*^{6/5}}\!Y_{[0]}\, \\
\mbox{}\!\!\!\!\!\!  t \!\!&\!\equiv\!\!\! &[\alpha_*\beta_*\kappa_*]^{1/5}s,
 \end{eqnarray}
 \end{subequations}
we obtain the normalized leading-order transition equations \cite{Ori:2000zn,Buonanno:2000ef,Kesden:2011ma}
\begin{subequations}\label{OTK}\begin{align}
  &  \left(\frac{dx_{[0]}}{dt}\right)^2 = -\frac{2}{3}x_{[0]}^3 - 2x_{[0]}t + y_{[0]},\\
 & \frac{d^2x_{[0]}}{dt^2} = -x_{[0]}^2 - t,\qquad   \frac{dy_{[0]}}{dt} = 2x_{[0]}.
\end{align}
\end{subequations}
The solution $x_{[0]}$ is the Painlev\'e transcendent of the first kind and $y_{[0]}$ is twice its first integral \cite{Compere:2019cqe}. We therefore proved that the transition equations \eqref{OTK} are unchanged in the presence of self-force.

In order to match with the inspiral as described below we consider the boundary condition as $t\rightarrow -\infty$ \cite{Ori:2000zn}
\begin{equation}
x_{[0]}\!=\! \sqrt{-t}+O(t^{-2}),\quad y_{[0]}\!=\!-\frac{4}{3}(-t)^{3/2} + O(t^{-1}). \label{BC}
\end{equation}
The Painlev\'e transcendent of the first kind is then uniquely defined from $t=-\infty$ to a finite $t \approx 3.41$ where $x_{[0]}\rightarrow -\infty$.

The transition equations are valid for all $s$ such that $\eta^{2/5}R_{[0]}(s) \ll 1$, $\eta^{6/5}Y_{[0]}(s) \ll 1$, and $\eta^{4/5}\xi_{[0]}(s) \ll 1$. At early times, using Eqs. \eqref{BC} and \eqref{xytDef}, we deduce that the transition regime breaks down when $\tau_\text{break}^{(-)} - \tau_*\sim -\eta^{-1}$. At late times, the transition equations break down at $s \sim s_{\text{break}}^{(+)}\sim\eta^0$ or, equivalently, at $\tau_{\text{break}}^{(+)}-\tau_* \sim \eta^{-1/5} \ll \eta^{-1}$. The range of validity of the transition equations is therefore 
\begin{equation}
-\eta^{-1} \ll \tau-\tau_* \ll \eta^{-1/5}. \label{rangetr}
\end{equation}

%The transition motion can therefore be defined from $s=-\infty$ to $s=s_{\text{merger}}>0$ as announced.

\section{Inspiral-transition matching}

The adiabatic inspiral has as range of validity ${\tau<\tau_*}$ where the bound arises because the expansion becomes singular at the LSO. The LSO is approached when ${\tau_*-\tau \ll\eta^{-1}}$ where $\eta^{-1}$ is the radiation reaction timescale. The transition solution is valid in the range \eqref{rangetr} and approaches the inspiral at early proper times with respect to the transition timescale $\tau_*- \tau \gg \eta^{-1/5}$. The overlapping region between the inspiral and the transition solution is 
\begin{equation}
-\eta^{-1} \ll \tau -\tau_* \ll -\eta^{-1/5}   
\end{equation}
with $\tau < \tau_*$ which is indeed a subset of \eqref{rangetr}. We will now match the transition solution as $s\!\rightarrow\!-\infty$ with the inspiral solution as $\tau\!\rightarrow\!\tau_*$ in the overlapping region. 

We consider the boundary condition \eqref{BC}. From Eqs. \eqref{X} and \eqref{fluxeq}, we have $\xi_{[1]}(s)=0$ and $\xi_{[2]}(s)=\lambda_* (-s)^{3/2}+O(s^{-1})$, where 
\begin{equation}\label{lambda}
\lambda_* \equiv -\frac{2}{3\eta}   \sqrt{\frac{\beta_*\kappa_*}{\alpha_*}} \left. \frac{\partial f_{\phi}}{\partial r}\right\vert_{[0]*}. 
\end{equation} 
Substituting in Eq. \eqref{exptau}, the dependence in $\eta$ and $  \tau$ recombines into a dependence in $\tilde \tau = \eta  \tau$ as $s\rightarrow -\infty$ as 
\begin{align}
\!\!r(s) &=   \sqrt{\frac{\beta_*\kappa_*}{\alpha_*}} \sqrt{\tilde \tau_*-\tilde \tau} +  r_{[0]}^*\nonumber +\eta^{2/5}O(s^{-2}) \\
& +O_s(\eta^{3/5}), \\
\!\! \ell(s) &=\lambda_* (\tilde\tau_*- \tilde \tau)^{3/2} \!+\!\kappa_* (\tilde \tau_*-\tilde \tau) \!+\! \ell^*_{[0]}+\eta^{6/5}O(s^{-1}) \nonumber \\
& +O_s(\eta^{7/5}).
\end{align}
This behavior asymptotically matches with Eqs. \eqref{inspiralexplicit} and \eqref{inspiralexplicit2} upon identifying
\begin{equation}
\kappa^*_{(0),2} = \kappa_* ,\;\; r_{(0),1}^*=\sqrt{\frac{\beta_*\kappa_*}{\alpha_*}},\;\; \kappa^*_{(0),3}= \lambda_*.\label{eqal}
\end{equation}
The three free parameters of the inspiral, namely $\kappa^*_{(0),2}$, $r_{(0),1}^*$, and $\kappa^*_{(0),3}$ are now fixed through the matching in terms of the parameters of the transition motion defined in Eqs. \eqref{kappa}, \eqref{defepszeta}, \eqref{defcoefs}, and \eqref{lambda}.

The deviation $\delta$ from quasi-circularity in the transition  \eqref{delta0tr} exactly matches at leading order with the deviation from quasi-circularity in the inspiral \eqref{inspiralexp} and \eqref{delta0in} thanks to the equality of coefficients $\pi_* \!\!=\!\! 3\sigma r_{(0)*}  \Omega_{(0)*} \Delta_*E_*/(2A_*)$. We also obtain as $s \rightarrow -\infty$
\begin{eqnarray}
    e(s) &=&  e^*_{[0],3} (\tilde\tau_*- \tilde \tau)^{3/2}+\kappa_* \Omega_{[0]*}(\tilde\tau_*- \tilde \tau) + e_{[0]*} \nonumber\\
  &  &+\eta^{6/5}O(s^{-1}) +O_s(\eta^{7/5}); \\ 
 e^*_{[0],3} \Omega_{[0]*}^{-1}&=&\lambda_*- \frac{4\alpha_*}{3\gamma_*} (r_{(0),1}^*)^3.\label{e03match}
\end{eqnarray}
We can also identify Eq. \eqref{e03match} with Eq. \eqref{cons3} after recognizing equivalent formulas 
\begin{eqnarray*}
\alpha_*= \frac{9\sigma \Delta_* \Omega_{(0)*} E_*}{4 r_{(0)*}^{1/2}A_*^{3}}, \; \beta_*=\frac{2\sqrt{A_*}\Omega_{(0)*}}{r_{(0)*}^{5/2}},\; \gamma_*=\frac{2\sigma \Delta_*}{r_{(0)*}^2 \sqrt{A_*}}.\nonumber %\label{coefm}
\end{eqnarray*} 
Summing up the formulas \eqref{frval},\eqref{delta0tr},\eqref{defepszeta}, the leading self-force in the inspiral near the LSO takes the final form
\begin{equation}\label{finalfr}
%\mbox{}\!\!\!\!\!\!\!\!f^r_{(0)} &\!\!=\!\!& \kappa_* \frac{\beta_* \epsilon_* - \alpha_* \zeta_*}{\alpha_*-\epsilon_*} (\tilde \tau_*-\tilde\tau)+O_{\eta}(\tilde \tau_*-\tilde\tau)^{3/2}, \\
f^r_{(0)} =O_{\eta}(\tilde \tau_*-\tilde\tau)^{3/2},
\end{equation}
which matches $\delta_{(0)}=O_{\eta}(\tilde \tau_*-\tilde\tau)^{3/2}$ in the inspiral motion. This completes this leading order matching.  The adiabatic inspiral equations are consistent with the assumption $\delta_{(0)}=0$, which implies from Eq. \eqref{deffr} that $f^r_{(0)}=0$ in Eq. \eqref{frin}. Eq. \eqref{r0inspiral} then becomes trivial and Eq. \eqref{dinspiral} becomes the radial evolution equation. Moreover Eq. \eqref{cons2} fixes $r_{(0),1}^*$ in terms of $\kappa_{(0),2}^*$. 

\vspace{-5pt}

\section{Conclusion}

We obtained the first exact consistent match of the adiabatic quasi-circular inspiral with the transition solution at leading order in the small mass ratio expansion. We proved that the leading-order transition solution including all self-force effects is determined in terms of the Painlev\'e transcendent of the first kind. This consolidates previous partial analyses for equal \cite{Buonanno:2000ef,Buonanno:2005xu} and small mass ratios \cite{Ori:2000zn,Kesden:2011ma,Compere:2019cqe,Burke:2019yek}. 

We proved that the adiabatic inspiral needs to take into account the secular change of angular velocity induced by radiation reaction in order to match the transition solution, consistently with the 2.5 post-Newtonian radiation reaction effect occuring in the post-Newtonian--post-Minkowskian formalism \cite{PhysRevLett.70.113,Blanchet:2013haa}. 

This mathematically self-consistent inspiral-transition motion in the small mass ratio expansion, once extended to higher orders and nonperturbatively resummed, would provide a new tool to further calibrate EOB waveforms \cite{Buonanno:2000ef,Buonanno:2005xu,Damour:2007xr,Damour:2009kr,Pan:2013rra,Nagar:2006xv,Bernuzzi:2010ty,Bernuzzi:2010xj,Bernuzzi:2011aj} using self-force theory. 

\vspace{4pt}
{\noindent \bf Acknowledgments.} We thank L. Blanchet, S. Gralla, and A. Pound and the anonymous referees for their very useful comments on the manuscript. G.C. is Senior Research Associate of the F.R.S.-FNRS and acknowledges support from the FNRS research credit J.0036.20F, bilateral Czech convention PINT-Bilat-M/PGY R.M005.19 and the IISN convention 4.4503.15. L.K. acknowledges support from the ESA Prodex experiment arrangement 4000129178 for the LISA gravitational wave observatory Cosmic Vision L3. 
\vspace{-10pt}

%\bibliography{refs}
%\bibliographystyle{utphys}

\providecommand{\href}[2]{#2}\begingroup\raggedright\endgroup

\end{document}